\documentclass[aps,prl,twocolumn,groupedaddress,nofootinbib,preprintnumbers]{revtex4}
\usepackage{graphicx}
\usepackage{dcolumn}
\usepackage{bm}
\usepackage{latexsym}
\usepackage{amsfonts}
\usepackage{amssymb}
\usepackage{amsmath}
\usepackage{verbatim}
\usepackage{color}
\usepackage{amsmath,amssymb}

\newcommand{\calR}{{\cal R}}

\begin{document}

\preprint{YITP-15-111}

\title{Inflationary Magnetogenesis with Broken Local $U(1)$ Symmetry}

\author{Guillem Dom\`enech}
\affiliation{Yukawa Institute for Theoretical Physics, Kyoto University}
\author{Chunshan Lin}
\affiliation{Yukawa Institute for Theoretical Physics, Kyoto University}
\author{Misao Sasaki}
\affiliation{Yukawa Institute for Theoretical Physics, Kyoto University}


\begin{abstract}
We point out that a successful inflationary magnetogenesis could be 
realised if we break the local $U(1)$ gauge symmetry during inflation. 
The effective electric charge is fixed as a fundamental constant, 
which allows us to obtain an almost scale invariant magnetic 
spectrum avoiding both the strong coupling and back reaction problems. 
We examine the corrections to the primordial curvature perturbation
due to these stochastic electromagnetic fields and find that, at both 
linear and non-linear orders, the contributions from the electromagnetic field
are negligible compared to those created from vacuum fluctuations.
Finally, the U(1) gauge symmetry is restored at the end of inflation.
\end{abstract}

\maketitle

{\bf Introduction} Magnetic fields are present throughout the universe and
 play an important role in many astrophysical process such as galaxy dynamics, 
pulsars, white dwarfs, and even black holes. However, its origin is still 
not well understood. Interestingly, in 2010, it was 
found by blazar observations that magnetic fields are present even 
in inter-galactic void regions, with the coherence length of a few Mpc
and the field strength larger than 
$10^{-15}G$~\cite{mag2010,mag20102,mag20103,mag2011}. 
The origin of such large scale cosmic magnetic fields, has been a 
challenging problem for years. It was suggested that inflation could be 
a prime candidate for the production of large scale magnetic fields,
provided the conformal invariance of the $U(1)$ gauge theory was broken 
during inflation \cite{turner1988}. 
For example, this is easily the case in higher dimensions, where the $U(1)$ 
gauge field is not conformally invariant.
 Many models are proposed so far but none of them is convincing enough,
 see \cite{Kandus:2010nw,Durrer:2013pga,Subramanian:2015lua} for recent reviews.

On the other hand, stochastic primordial magnetic fields might also leave 
many imprints on CMB, such as spectral distortions~\cite{Puy:1998sv,Jedamzik:1999bm},
modifications of temperature and polarization spectra, 
Faraday rotation of CMB polarization~\cite{Kosowsky:2004zh,Yadav2012,Bonvin:2013tba},
and so on.  The most recent observational constraint from CMB 
is roughly $B_{\text{Mpc}}<10^{-9}G$ by Planck~\cite{Ade:2015cva}.

One of the simplest, gauge invariant and well studied model 
is described by the Lagrangian \cite{turner1988,ratra1992}
\begin{eqnarray}
\mathcal{L}_{EM}=-\frac{1}{4}f^2(\phi)F_{\mu\nu}F^{\mu\nu},
\end{eqnarray}
where $F_{\mu\nu}\equiv \partial_{\mu}A_{\nu}-\partial_{\nu}A_{\mu}$. 
This model is equivalent to Bekenstein's variable charge 
theory \cite{Bekenstein1982}. Unfortunately, to produce large enough 
amplitude of magnetic field at large scale, this model suffers from 
either back reaction or strong coupling problem \cite{Demozzi:2009fu}. 
Furthermore, the primordial stochastic electric and magnetic fields 
behave as isocurvature perturbations, and hence they may also contribute 
to the CMB temperature anisotropies at both linear and nonlinear orders.
Requiring the absence of the back reaction and strong coupling problems, 
as well as the absence of too large isocurvature perturbations from the 
eletromagnetic field, yields that inflation must happen at an extremely low energy 
scale \cite{Suyama:2012wh,Fujita:2012rb,Fujita:2013pgp,Fujita:2014sna,Ferreira:2014hma,Green:2015fss}.
For instance, one of the most stringent lower bounds was found 
in \cite{Ferreira:2014hma}, in which only a very fine tuned scenario at
 an energy scale of inflation as low as $10^{-2}GeV$ can explain 
the observed void magnetic field today. Basically it either rules out almost 
all models of inflation or rules out the inflationary origin 
of large scale magnetic fields.

{\bf Magnetogenesis with broken $U(1)$ symmetry~~} In this letter, 
we point out that all of the above problems could be avoided if the 
local $U(1)$ gauge symmetry is broken during inflation. In contrast 
to Bekenstein's variable charge theory \cite{Bekenstein1982}, we demote 
the electric charge as a fundamental coupling constant by considering 
the following modified QED Lagrangian during inflation,
\begin{eqnarray}\label{modQED}
\mathcal{L}=&&
-\frac{1}{4}f^2(\phi)F_{\mu\nu}F^{\mu\nu}
- e_0f(\phi)A_{\mu}\bar{\psi}\gamma^{\mu}\psi\nonumber\\
&&+i\bar{\psi}\gamma^{\mu}
\left(\partial_\mu+\Gamma_\mu\right)\psi-m\bar{\psi}\psi,
\end{eqnarray}
where $e_0$ is the value of the electric charge today, 
$\psi$ is a charged fermion and $\Gamma_\mu$ is the spin connection.
Thus, the standard local $U(1)$ gauge symmetry in QED is certainly 
broken. The residual symmetries are global symmetries
\begin{eqnarray}
\psi&\to&\psi e^{i\alpha},\nonumber\\
A_{\mu}&\to& \lambda A_{\mu},\nonumber\\
f(\phi)&\to& \lambda^{-1}f(\phi),
\end{eqnarray}
as well as the on-shell local symmetry
\begin{eqnarray}
\label{osgauge}
A_{\mu}\to A_{\mu}+\partial_{\mu}\sigma.
\end{eqnarray}
A possible origin for the effective action (\ref{modQED}) is a 
braneworld scenario, in which charged matter fields live on the brane 
and the gauge field lives in the bulk, though the actual
realization seems rather non-trivial.

We also mention that the above on-shell gauge symmetry resembles the
case of the on-shell diffeomorphism invariance in Einstein gravity.
In contrast to the off-shell gauge symmetry of standard QED, 
the action is invariant under the above gauge transformation only 
when the equation of motion is imposed. The equation of motion of the
gauge field reads
\begin{eqnarray}
\nabla_{\mu}(f^2F^{\mu\nu})-e_0f\bar{\psi}\gamma^{\nu}\psi=0.
\end{eqnarray}
Taking the divergence of the above equation of motion,
by noting that $\nabla_{\mu}\nabla_{\nu}(f^2F^{\mu\nu})\equiv0$, 
we must have
\begin{eqnarray}
\label{chargecon}
\nabla_{\mu}\left(e_0f\bar{\psi}\gamma^{\mu}\psi\right)=0.
\end{eqnarray}
Under the gauge transformation (\ref{osgauge}), the variation of the action reads
\begin{align}
\delta_g S&=-\int\sqrt{-g}\, e_0f\partial_{\mu}\sigma\bar{\psi}\gamma^{\mu}\psi
 \nonumber\\
=&-\int\sqrt{-g}
\left[\nabla_{\mu}\left(e_0f\sigma\bar{\psi}\gamma^{\mu}\psi\right)
-\sigma\nabla_{\mu}\left(e_0f\bar{\psi}\gamma^{\mu}\psi\right)\right],
\end{align}
where $\delta_g$ denotes the variation $\delta_g A_\mu=\partial_\mu\sigma$.
The first term is just the total derivative, and the second term 
gives the generalised current conservation equation (\ref{chargecon}), 
and thus we have $\delta_gS=0$. Similar to the case of GR,
 where the gauge symmetry is also on-shell symmetry, the anomaly 
would appear at the quantum level and spoil the gauge
 symmetry (\ref{osgauge}). However, by imposing a more fundamental 
symmetry (such as SUSY), similar to what people did in supergravity,  
in principle all loop corrections can be cancelled out and thus the 
anomaly may be removed. But this is a topic beyond the scope of the 
present paper.

Another way to express our model is to do the scaling 
$A_{\mu}\to f^{-1}A_{\mu}$. Our action is reduced to 
\begin{eqnarray}\label{modqed}
\mathcal{L}=&&-\frac{1}{4}F_{\mu\nu}F^{\mu\nu}
+i\bar{\psi}\gamma^{\mu}D_{\mu}\psi-m\bar{\psi}\psi\nonumber\\
&&-F_{\mu\nu}Q^{\mu}A^{\nu}-\frac{1}{2}Q_{\mu}Q^{\mu}A_{\nu}A^{\nu}
+\frac{1}{2}\left(Q_{\mu}A^{\mu}\right)^2,
\end{eqnarray}
where $Q_{\mu}\equiv-\partial_{\mu}\left(\ln f\right)$ 
and $D_\mu\equiv\partial_\mu+ie_0A_\mu+\Gamma_\mu$.
 Our theory is the standard QED but the gauge symmetry is explicitly broken 
by the scalar field $\phi$. A related example was studied in \cite{Liu:2015tza}\cite{Tasinato:2014fia}.
Note that our model is free from the strong coupling problem because 
we have fixed the electric charge as a fundamental constant, as clear from
the above form of the Lagrangian.  

The global symmetry between $A_{\mu}$ and $f(\phi)$ is also broken if we
 introduce a potential to the inflaton $\phi$ (which may arise from the 
SUSY breaking), to generate the well-behaved kinetic coupling 
function $f(\phi)$. It is known that we can obtain a scale-invariant 
magnetic spectrum if $f^2\sim a^4$ \cite{Bamba:2006ga,Demozzi:2009fu}. 
However, in this case if we integrate over the whole momentum space, 
the energy density of the magnetic field would diverge
logarithmically~\cite{Barnaby:2012tk}. The divergence
 becomes even much worse if we take into account the slow-roll correction. 
Thus we relax the assumption on the coupling function as 
\begin{eqnarray}\label{f2}
f^2\sim a^4\cdot e^{(4\epsilon +\nu)N},
\end{eqnarray}
where $\epsilon\equiv -\dot{H}/H^2$ is the slow-roll parameter
during inflation, which is assumed to be constant for simplicity, 
$\nu$ is a new parameter and we require $\nu\ll1$ to generate a nearly 
scale-invariant magnetic spectrum. $N$ is the number of $e$-folds
counted backward from the end of inflation, thus $N=0$ at the end of 
inflation.

In the momentum space, the action of the gauge field reads
\begin{align}
S&=
\frac{1}{2}\sum_s\int d\tau\int d^3k f^2(\phi)
\left[(A_k^{s})'(A_{-k}^{s})'-k^2A_k^{s}A_{-k}^{s}\right],
\end{align}
where $s$ is the polarization index, $\tau$ is the conformal time,
and the prime ($~'$) denotes the derivative w.r.t. the conformal time. 
The canonical conjugate momentum of the gauge field is defined by 
\begin{eqnarray}
\pi_k=\frac{\delta S}{\delta A_k'}=f^2A_{-k}'.
\end{eqnarray}
where we have omitted the polarization index for notational simplicity.
We quantize our system by imposing the following canonical
commutation relation,
\begin{eqnarray}\label{commu}
\left[A_{k_1},\pi_{k_2}\right]=i\delta(\textbf{k}_1-\textbf{k}_2),
\end{eqnarray}
 In terms of the creation and annihilation operators, 
the gauge field is expanded as 
\begin{eqnarray}
A_k= u_k a_k+u_k^*a_{-k}^\dagger, 
\end{eqnarray}
where, due to the commutation relation (\ref{commu}), the mode function $u_k$ 
is normalised as 
\begin{eqnarray}\label{noma}
u_ku_k^*{}'-u_k^*u_k{}'=\frac{i}{f^2}.
\end{eqnarray}
Taking the variation of the action w.r.t the gauge field, 
the equation of motion reads
\begin{eqnarray}
u_k''+\frac{2f'}{f}u_k'+k^2u_k=0.
\end{eqnarray}
At the subhorizon limit $k\tau\to-\infty$, the spacetime is asymptotically 
Minkowskian, and the term $\frac{2f'}{f}$ is much smaller than $k$.
Thus we may assume the standard Minkowski vacuum, and 
the mode function can be given by the WKB solution,
\begin{eqnarray}
u_k=\frac{1}{f\sqrt{2k}}e^{-ik\tau},
\end{eqnarray}
where the pre-factor $1/f\sqrt{2k}$ is fixed by the Klein-Gordon
normalisation condition (\ref{noma}).
Due to the nearly exponential expansion of the background, the physical 
scale eventually exceeds the Hubble horizon and since $f'/f=O(a'/a)$,
we can neglect the $k^2$ term in the superhorizon limit $k\tau\to 0$. 
In this limit, the solution is
\begin{eqnarray}
u_k=c_1+c_2\int_0^\tau \frac{d\tau}{f^2},
\end{eqnarray}
where $c_1$ represents the amplitude of the constant mode, and $c_2$ the 
decaying mode. Given the coupling function (\ref{f2}), it implies that 
the energy density of the electric field, which is dominated by the
decaying mode, is much smaller than that of the magnetic field.
Hereafter we ignore the decaying mode $c_2$ and hence the electric field.
The $c_1$ may be computed by matching the subhorizon and superhorizon solutions
at horizon crossing,
\begin{eqnarray}
c_1\simeq \frac{1}{f_k\sqrt{2k}},
\end{eqnarray}
where $f_k$ is the coupling function at horizon crossing time
of the wavenumber $k$. 
Thus the power spectrum of the magnetic field can be calculated as 
\begin{eqnarray}
P_B(k)=\frac{k^5f^2}{\pi^2 a^4}|u_k|^2
=\frac{k^4}{2\pi^2a^4}\cdot\frac{f^2}{f_k^2}.
\end{eqnarray}
At horizon crossing, we have $k=a_*H_*=a_*H_fe^{\epsilon N}$, 
where $H_f$ is the Hubble constant at the end of inflation.
Setting
\begin{align}
f^2(N)=\exp[-4N+(4\epsilon+\nu)N]\,; \quad N>0\,,
\end{align}
and $f^2=1$ at and after the end of inflation,
the power spectrum is found as
\begin{eqnarray}
P_B(k;N)=\frac{H(N)^4}{2\pi^2}e^{\nu(N-N_k)}.
\end{eqnarray}
We assume that $\nu>0$ so that we can have a convergent energy density 
of the magnetic field. The energy density of the magnetic field is 
given by integrating over the whole momentum space (with UV cutoff at 
$k=aH\equiv k_*(N)$),
\begin{eqnarray}
\rho_B(N)&=&\frac{1}{2}\int_0^{k_*(N)} \frac{dk}{k}P_B
\nonumber\\
&=&\frac{H^4(N)}{4\pi^2}e^{\nu N}\int_N^\infty dN_k e^{-\nu N_k}
\nonumber\\
&=&\frac{H^4(N)}{4\pi^2\nu}.
\end{eqnarray}
By requiring that the energy density of the magnetic field is much 
smaller than background energy density, we have a constraint on 
the parameter $\nu$,
\begin{eqnarray}
\frac{H_f^2}{M_p^2}\ll\nu\ll1,
\end{eqnarray}
where the upper bound is set to ensure a nearly scale-invariant magnetic spectrum, and $H_f$ is the Hubble constant at the end of inflation.
We see that this is actually a very week constraint. 

The current magnetic strength could be estimated as follows. 
Assume that inflation happened at GUT scale, $H_f\sim 10^{-6}M_p$, we have 
\begin{eqnarray}
B\sim H_f^2\nu^{-1/2}\sim \nu^{-1/2}10^{-12}M_p^2\sim \nu^{-1/2}10^{46}G
\end{eqnarray}
at the end of inflation.  Assuming that all the energy of the inflaton 
is transfered to radiation at once at the end of inflation, we have 
\begin{eqnarray}
T^4_{\text{CMB}}\sim M_p^2H_f^2\left(a_f/a_0\right)^4,
\end{eqnarray}
where $T_{\text{CMB}}$ is the CMB temperature today,
$a_f$ is the scale factor at the end of inflation, and $a_0$ is the 
scale factor today.
The energy density of the magnetic field evolves also as radiation,
$\rho_B\sim a^{-4}$, and its strength today reads
\begin{eqnarray}
B_0\sim B \left(\frac{a_f}{a_0}\right)^2
\sim B\times\frac{T_{\text{CMB}}^2}{M_pH}\sim \nu^{-1/2}10^{-12} G,
\end{eqnarray}
which is sufficiently large to explain the large scale magnetic field in the void.
 The most recent observational constraint from CMB is roughly
 $B_{\text{Mpc}}<10^{-9}G$ by Planck~\cite{Ade:2015cva}, 
and it translates to the constraint, 
\begin{eqnarray}
\label{cstnu}
10^{-6}<\nu\ll1.
\end{eqnarray}

Let's end this section with a remark. We assume that after inflation,
the inflaton $\phi$ is trapped at the bottom of potential and 
we recover the standard $U(1)$ gauge theory. In the radiation dominant phase, the presence of the high conductivity would only dissipate the electric field, and keep the magnetic field frozen at super horizon scale.

{\bf The corrections to the primordial perturbations} 
The primordial curvature perturbation also receives contributions from
the electromagnetic field. The contributions to both linear and 
non-linear perturbations may be estimated by the $\delta N$ 
formalism \cite{deltaN1,deltaN2,deltaN3,deltaN4,Lyth:2005fi}. 
The $\delta N$ formalism is a very powerful tool for the understanding
of the perturbative physics on cosmological scales. It is essentially equivalent
to focusing on the leading order terms in spatial gradient expansion,
called the separate universe approach.
According to it, the evolution of the local Hubble patch is well-approximated 
by the evolution of an unperturbed universe. 
The curvature perturbation at $t=t_f$ is given by the perturbation in
the number of $e$-folds between the initial flat slice at $t=t_i$
and a final comoving slice at $t=t_f$ when the universe is in 
the adiabatic limit,
\begin{eqnarray}
\calR_c(t_f,\textbf{x})=\delta N\equiv N(t_i\to t_f,\textbf{x})-N_0(t_i\to t_f),
\end{eqnarray}
where $N_0\equiv \ln\left[a(t_f)/a(t_i)\right]$ is the 
unperturbed number of $e$-folds. In our case, the inflationary 
expansion history is parameterized by the value of the scalar field $\phi$,
\begin{eqnarray}
N(\phi)=\int_{\phi}^{\phi_f}\frac{H(\tilde{\phi})}{\dot{\tilde{\phi}}}d\tilde{\phi}
\simeq\int_{\phi_f}^{\phi}\frac{\rho_{\text{inf}}(\tilde{\phi})
+\rho_B(\tilde{\phi})}{M_p^2V_\phi(\tilde{\phi})}d\tilde{\phi},
\end{eqnarray}
where $\rho_{\text{inf}}$ is the energy density of the inflaton, 
$V_\phi\equiv\partial V/\partial\phi$, 
and we have used the Friedmann equation, the equation of motion for $\phi$, 
and the slow-roll approximation.
 According to the $\delta N$ formalism, we have 
\begin{eqnarray}
\calR_c(t_f,\textbf{x})=\frac{\partial N}{\partial \phi}\delta\phi
=\frac{\rho_{\text{inf}}(\phi)\delta\phi}{V_\phi(\phi)}
+\frac{\rho_B(\phi)\delta\phi}{V_\phi(\phi)},
\label{calRc}
\end{eqnarray}
where $\delta\phi$ is to be evaluated on the initial flat slice at $t=t_i$.
To evaluate the curvature perturbation for a given $k$, we simply identify
$t_i$ with the horizon crossing time determined by $a(t_k)H(t_k)=k$.
The first term in (\ref{calRc}) is due to the inflaton density fluctuation, 
and the second term is the additional contribution from the magnetic field. 
Given the constraint on the parameter $\nu$, (\ref{cstnu}), 
the contribution from the magnetic field is completely negligible. 
For instance, for parameter $\nu\sim \mathcal{O}(10^{-2})$, we have 
$\rho_B/\rho_{\text{inf}}\sim 10^{-12}$. 

The same conclusion applies to the non-linear perturbations. 
For the bi-spectrum (3-point function), the size of non-Gaussianity 
is characterised by the non-linear parameter
\begin{eqnarray}
-\frac{3}{5}f_{NL}=\frac{1}{2}
\frac{\partial^2N/\partial\phi^2}{\left(\partial N/\partial\phi\right)^2}
\supset\frac{3H\dot{\rho}_B-\rho_BV''}{\rho^2_{\text{inf}}}+...
\end{eqnarray}
Taking into account that the energy density of the magnetic field is 
much smaller than that of the inflaton field, and it evolves very slowly,
the contribution to the non-Gaussianity from the magnetic field is 
also completely negligible. The same conclusion can be trivially 
extended to the tri-spectrum (4-point function). 

The magnetic field may also source tensor 
perturbations~\cite{Ferreira:2014hma,Barnaby:2012tk}. The transverse and
 traceless part of the electromagnetic energy-momentum tensor
appears as a source term on the right-hand side of the equation of motion 
for the tensor perturbation,
 \begin{eqnarray}
 \frac{1}{2a^2}\left(\gamma_{ij}''+\frac{2a'}{a}\gamma_{ij}'
+k^2\gamma_{ij}\right)\simeq -\frac{1}{M_p^{2}}\left(B_iB_j\right)^{TT}.
 \end{eqnarray}
As an order of magnitude estimation of the contribution from the
electromagnetic field, we set the left-hand to be $H^2\gamma_{ij}$
and the right-hand side to be $\frac{H^4}{M_p^2\nu}$. Thus we obtain
\begin{eqnarray}
\gamma_{ij}^{\text{em}}\sim 
\frac{H^2}{M_p^2\nu}\ll \gamma_{ij}^{\text{vac}}\sim\frac{H}{M_p},
\end{eqnarray}
where the middle double inequality holds because $\nu\gg H/M_p\lesssim 10^{-6}$.
Thus the vacuum fluctuation always dominates over 
the one generated by the electromagnetic field. 

{\bf Conclusion and discussion} The theoretical explanation on the origin 
of the large scale magnetic field has been a challenging problem for years. 
In this letter, we have pointed out that if we break the local $U(1)$ symmetry
by promoting the effective electric charge to be a fundamental constant,
 the sufficient magnitude of stochastic magnetic fields may be generated 
during inflation without encountering neither the back reaction problem 
nor the strong coupling problem. 
The local $U(1)$ symmetry is restored at the end of inflation.
We have examined that the contributions
from the generated electromagnetic field to the primordial curvature perturbation
are completely negligible both at linear and non-linear orders.

An additional interesting fact is that a broken $U(1)$ symmetry may 
have a more significant consequence such as baryon asymmetry. 
The broken $U(1)$ symmetry may lead
 to the lepton number violation. The lepton number violation translates to 
the baryon number violation through the weak interaction which meditated by sphaelerons 
\cite{leptogenesis}. Thus we may create the baryon asymmetry
 by creating the lepton number during inflation. We hope to come back to
this issue in the future.
\\

\noindent
{\bf Acknowledgments}
This work was supported in part by MEXT KAKENHI Grant Number 15H05888. 
CL is supported by JSPS fellowship. We would like to thank H. Firouzjahi, A. E. Romano, T. Tanaka, A. Vikman, M. Werner for helpful discussions.

\end{document}